\documentclass[fleqn,10pt]{wlscirep}
\title{Entropy Measures of Human Communication Dynamics}

\author[1,*]{Marcin Kulisiewicz}
\author[1]{Przemysław Kazienko}
\author[2]{Bolesław K. Szymański}
\author[1]{Radosław Michalski}
\affil[1]{Wroclaw University of Science and Technology, Department of Computational Intelligence, Wroclaw, 50370, Poland}
\affil[2]{Rensselaer Polytechnic Institute, Department of Computer Science, Troy, NY 12180-3590, US}

\affil[*]{marcin.kulisiewicz@pwr.edu.pl}

\begin{abstract}
Human communication is commonly represented as a temporal social network, and evaluated in terms of its uniqueness. We propose a set of new entropy-based measures for human communication dynamics represented within the temporal social network as event sequences. Using real world datasets and random interaction series of different types we find that real human contact events always significantly differ from random ones. This human distinctiveness increases over time and by means of the proposed entropy measures, we can observe sociological processes that take place within dynamic communities.
\end{abstract}
\begin{document}

\flushbottom
\maketitle
%
%
\thispagestyle{empty}

\section*{Introduction}
Despite living in social communities and witnessing people communicate, at the first glance, we may not recognize clear patterns or trends of dynamic changes in communication -- the general impression my be that people interact almost randomly. Even though many studies\cite{zhao2011entropy, song2010limits, pham2013ebm, eagle2006reality} show that human interactions are not random, still some vital questions need to be addressed: how specific and how stable over time they are. Additionally, communication traces are the main source for interactions represented by social networks\cite{nicosia2013graph}, hence, the questions about communication dynamics simultaneously address the problem of stability of temporal networks. 

Despite the fact that temporal social networks have been studied for several years, there is no fixed and commonly agreed set of measures quantifying their dynamics. It is partially caused by the fact that there are many representations of temporal networks, such as event sequences, interval graphs, time windows, etc. It is hard to develop a comprehensive measure that will cover all the models. Therefore, we may expect that the development of dynamic measures will proceed differently than in the case of static networks. 


One of the most important concepts introduced in temporal setting is the time-respecting path, i.e. the path connecting nodes $v_i$ and $v_j$ in such a way that all intermediate nodes are visited in non-decreasing time order~\cite{grindrod2011communicability}. Starting with that metric, it was possible to define a number of natural subsequent measures, such as temporal connectedness~\cite{nicosia2013graph} between nodes representing the reachability from the source node to destination node in a given time, temporal diameter as a largest temporal distance between any two nodes or characteristic temporal path length that defines the temporal distance over all pairs of nodes~\cite{tang2010small}. Another important aspect of time-varying networks is the interevent time distribution~\cite{karsai2011small} that defines the frequency of events; it can be used to verify how bursty is the behavior in a given network. To quantify differences in burstiness, the expected number of short-time interactions is used  to characterize the early-time dynamics of a temporal network \cite{doyle2017effects}.
Lastly, a number of centrality measures were adapted or developed from scratch to describe the position of the node in the network, in particular: temporal betweenness~\cite{tang2010analysing}, temporal closeness~\cite{kim2012temporal}, and temporal degree~\cite{tang2010analysing}.

Entropy-based measures, in turn, were utilized by Takaguchi et al.\cite{takaguchi2011predictability} to evaluate the predictability of the partner sequence for individuals. In 2013, Kun Zhao et al.\cite{zhao2013models} proposed entropy-based measure to quantify how many typical configurations of social interactions can be expected at any given time, taking into account the history of the network dynamic processes.

We use the entropy to capture human communications dynamics -- event sequences (ES) depicting human interactions, which are also one of the basic lossless representations of temporal network\cite{holme2015modern}. In general, an event sequence is a time ordered list of interactions between pairs of individuals/agents within a given social group. 

Three main approaches to compute entropy for temporal networks represented as an event sequence are proposed: 1) the first order entropy, based on the probability of a node to appear as a speaker, or in other words, an initiator of event, 2) the second order entropy, based on probability of the event occurrence, that is probability of interaction between unique pair of nodes, 3) the third order entropy denoting probability of succession appearance, i.e. probability of unique pair of events. Each type of entropy captures different aspect of dynamics and have potential to be useful for different applications. For each new entropy measure, its maximum value can be estimated for a given number of nodes. This value is used for normalization and definition of relative entropy measures that allow us to compare entropies for different datasets.

This paper is organized as follow. In the first section, we present results of our experiments followed by main findings and conclusions. The second section broadly discusses meaning of findings and provides some insight for further work. The last section contains the detailed description of our experiments: experimental setup, datasets used and definitions of all entropies.


\section*{Results}
\begin{figure}[ht]
\centering
\includegraphics[width=\linewidth]{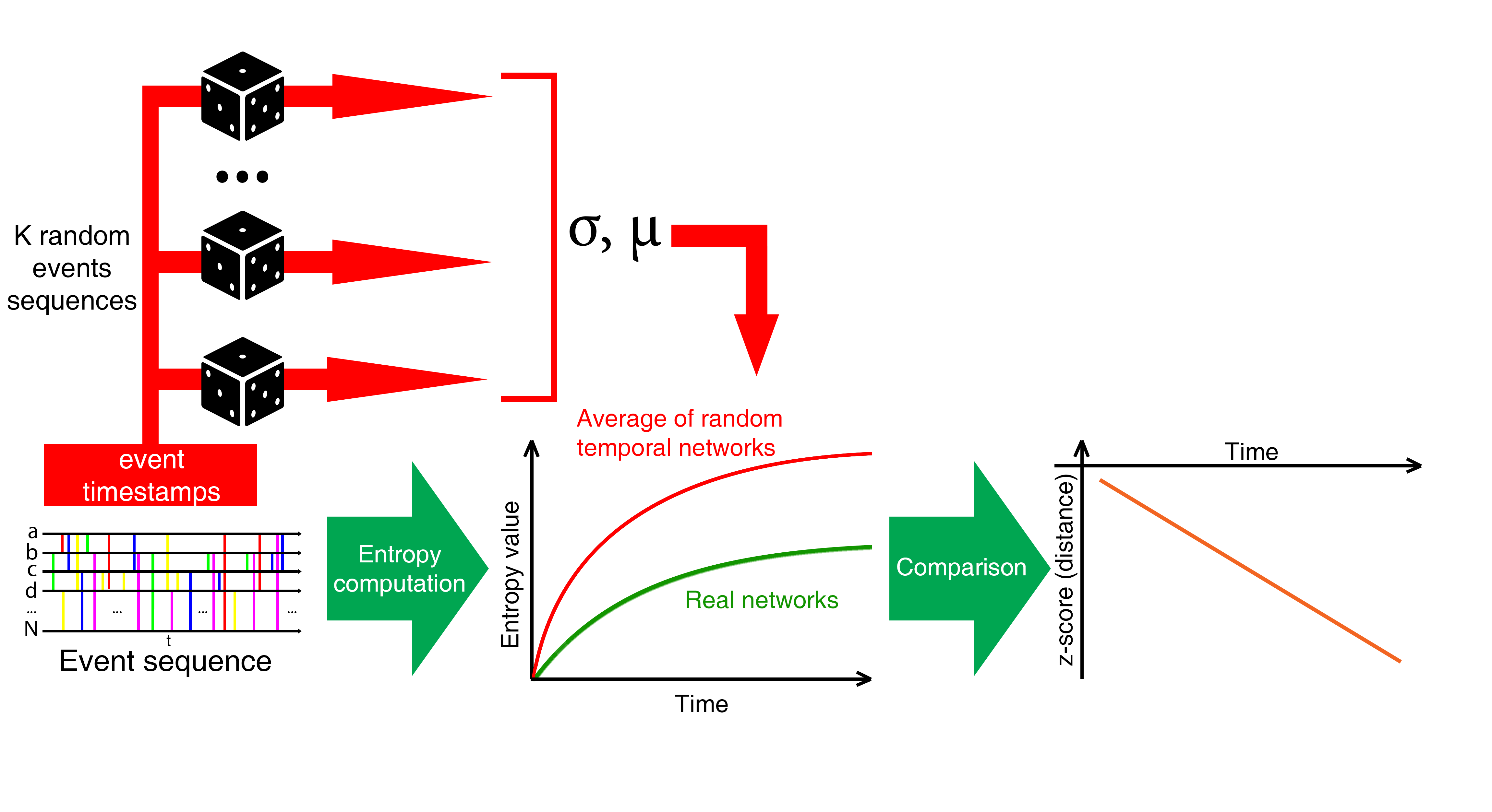}
\caption{General schema of experiments. K=100 was used. From original (real) event sequence, event timestamps are extracted as a base for random sequence generator. Entropy value is computed for real event sequence and artificial sequences. We compare results for real data with summarized results for artificial data using Z-score.}
\label{fig:schema}
\end{figure}

We compute entropy values for 4 different dataset with data of real human interactions: (1) face-to-face meetings at HyperText conference, (2) text messages exchanged between students for 6 semesters (NetSense), (3) email communications in the manufacturing company, and (4) face-to-face interactions between patients and hospital staff members. We compute time-line of entropy by taking a window from the beginning of network existence to point in time that we want to know the entropy value. In other words, we compute entropy cumulatively for on-line stream of interaction data. To provide the baseline for real event sequences, we generate 100 artificial event sequences for each dataset with the same numbers of nodes, events and timestamps by randomly reselecting pairs of nodes involved in each event. In the static networks, such procedure would be called rewiring. The average value of entropy for random event sequences is computed and compared against the values for the real network using Z-score -- the distance measure that, in general, shows the number of standard deviations by which the value of entropy for real sequence is above the mean value of random streams. The negative values of Z-score mean that entropies for real data are smaller than random ones and greater the difference is more negative Z-scores are. The general concept of experiments is presented in Fig. \ref{fig:schema}

\begin{figure}[ht]
\centering
\includegraphics[width=\linewidth]{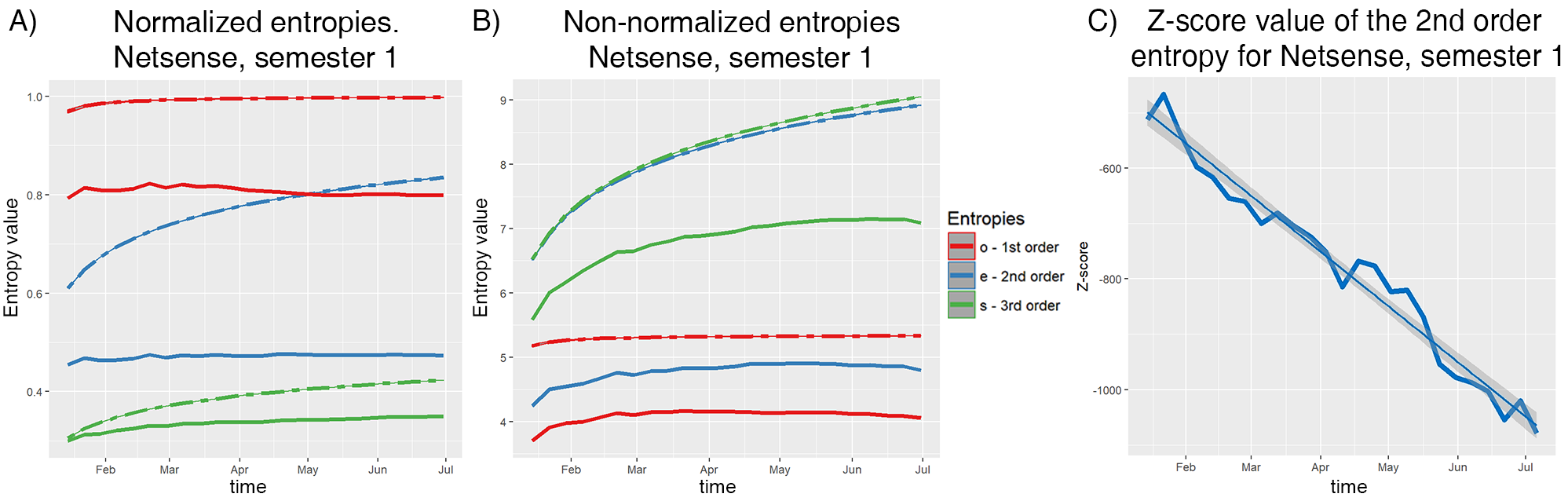}
\caption{The NetSense dataset, the 1st semester. A) Values of normalized entropies. Solid lines refer to the original event sequence and dashed ones present the average value for the baseline -- random sequences. B) Values of non-normalized entropies. C) Z-score for non-normalized second-order entropy with the computed trend and marked standard deviation (gray area).}
\label{fig:entr_comp_zscore}
\end{figure}

The first observation made about the nature of entropy is that the maximum value of entropy is non-decreasing over time since it directly depends on the non-decreasing number of distinct nodes in the event sequence. By normalizing entropy with its maximum, we obtain the relative entropies within the range [0,1]. Our experimental results show that entropy of random networks tend to reach the maximum value faster for first-order entropies and slower for higher-order ones. In Fig.\ref{fig:entr_comp_zscore}A, we can observe that entropy for random sequences have the shape suggesting that they converge to some maximum value, i.e. 1 in case of the normalized entropy. In Fig.\ref{fig:entr_comp_zscore}B we can observe similar tendency for non-normalized entropy. However, the relative entropy values for the real network seem to stabilize earlier around the smaller value. We can clearly observe such case for first-order entropy as well as converging shape for higher-order ones. The similar observations were made for all other examined datasets.

\begin{figure}[ht]
\centering
\includegraphics[width=\linewidth]{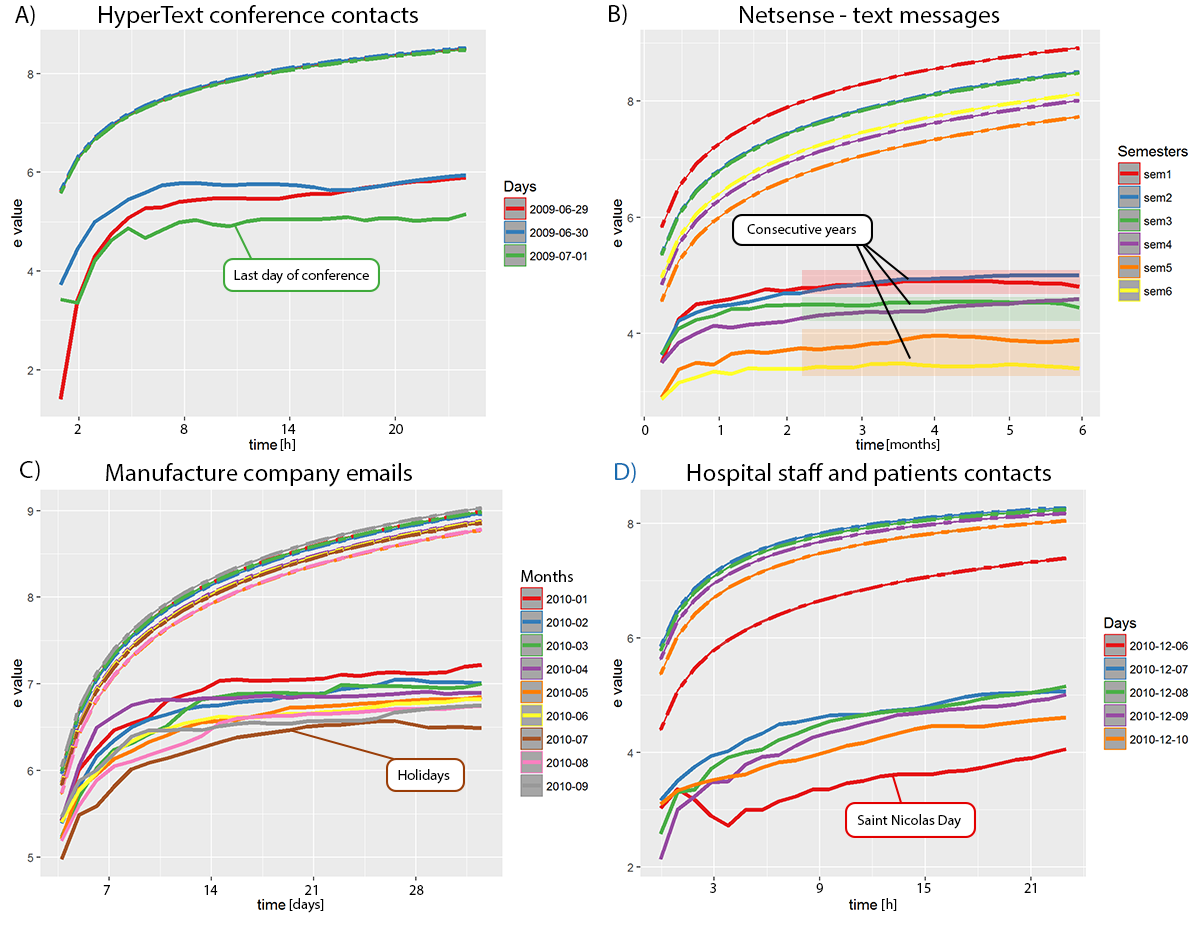}
\caption{Value of non-normalized second-order entropy for all examined datasets. Solid lines refer to the real event sequences; upper dashed lines -- to average values of random sequences. Each dataset is divided into parts for more convenient analysis. Parts were selected empirically. Different level of entropies for random sequences (especially for NetSense and hospital) comes from either smaller or greater number of interacting nodes in a given period. A) In consecutive days of conference entropy of communication decreases which is especially clear for last day of conference. B) Students tend to be more selective in their communication in later semesters than at the beginning of studies. C) Manufacture company employees communicate with similar dynamic over time but decreasing tendency of the entropy can be still observed with the exception of one month probably related to holiday period. D) Hospital staff and patients contacts shows decreasing entropy over consecutive days with the exception of 6th of December, usually celebrated as Saint Nicolas Day, which may influence contacts dynamic.} 
\label{fig:results}
\end{figure}

We split each dataset into reasonable parts selected empirically for more convenient analysis. Most clear observations were noted for non-normalized second-order entropy, see Fig.\ref{fig:results}, even though the same phenomena can be seen for all datasets and all entropies. The main finding that can be derived from our results is that entropy decreases over time except for some rare cases, which are explained later on. The results for face-to-face contacts on the first two days of the conference, see Fig. \ref{fig:results}A, are similar in terms of their dynamics, however, the last day is significantly different. It means that participants know each other much better on the last day and they interact much more consciously, i.e. with the smaller number of peers. A similar effect is observable for university students, see Fig.\ref{fig:results}B. The entropy decreases with each consecutive year of study and it is the lowest for the last, sixth semester. Further, the results of manufacturing company emails communication shows that for consecutive months value of entropy decreases with the exception of June 2010, see Fig.\ref{fig:results}C. We suppose that this month breaks from the pattern because of holidays - it may be the month when majority of employees go on vacation, what significantly changes dynamics of communication. Similarly, for face-to-face contacts among hospital stuff and patients, Fig.\ref{fig:results}D, we can note that entropy decreases in consecutive days except on the 6th of December. This day is usually celebrated as Saint Nicolas Day, which makes people significantly change their common pattern of communication.

We also measure distance of real sequence entropy from random sequence entropy using Z-score distance measure. The results confirm that there is a clear difference between reality and randomness. A sample plot of Z-score is presented in Fig.\ref{fig:entr_comp_zscore}C. For more results see Supplementary Fig.\ref{fig:zscore_multi} We can observe that Z-score decreases over time or in other words the difference between reality (smaller and stable over time entropy) and randomness (greater entropy and still growing in time) becomes more and more clear over time.

\begin{figure}[ht]
\centering
\includegraphics[width=\linewidth]{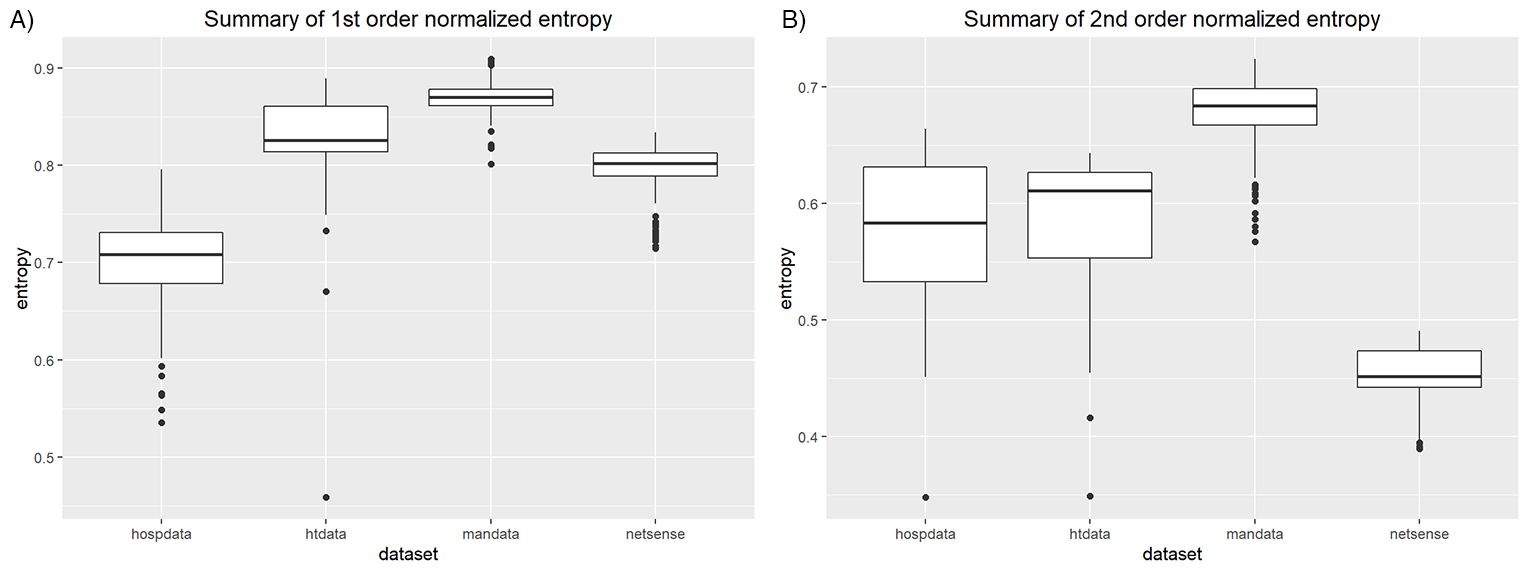}
\caption{Comparison of average value of normalized entropy: A) the first-order, B) the second-order. X-axis labels: \textit{hospdata} - dataset of hospital face-to-face contacts, \textit{htdata} - conference face-to-face contacts, \textit{mandata} - email communication in manufacturing company, \textit{netsense} - student text communication. Boxes shows median (black horizontal line) with the 1st and the 3rd quartile. Points refers to outliers and vertical lines to range of main observations.} 
\label{fig:dataset_comp}
\end{figure}

To show the difference between datasets, we compare entropy values, i.e. their normalized versions to exclude network size effect (different number of individuals), separately for the first- and second-order entropies, see Fig.\ref{fig:dataset_comp}. The greatest first-order value and lowest deviation is observed in the manufacturing company. It means that almost every employee needs to show up every working day in the company and interacts with the same frequency and stability of contacts with most of the other workers (the greatest second-order entropy). This suggests that communication in the company is decentralized and rather 'flat'. Patients in the hospital appear and disappear (low first-order entropy) but if they are present, they interact more randomly than students, who communicate much more within their encapsulated social/learning groups (low second-order entropy). Randomness of interactions between hospital staff members and patients as well as conference participants is comparable (second-order entropy) - they do not know each other so much, even though the first-order values suggest that there is less rotation among conference attendees appearance (first-order) than in hospital. The diversity of contacts (high standard deviation of second-order entropy) in hospital is the greatest, it means that depending on time, the social groups are more or less integrated, e.g. interactions among staff members and between patients are different. Interactions among students and employees are most stable (low standard deviation). Based on these observations, we conclude that different approaches to entropy computation (entropy order) can measure different aspects of communication dynamics.

\section*{Discussion}

The results of our experiments provide some interesting insights about human communication dynamics. Firstly, we can confirm the general intuition that people do not communicate randomly. This obvious fact now finds quantitative confirmation also in the temporal network context.

The second important observation is that entropy decreases over time, i.e. for consecutive periods. Referring to the examined dataset, we can explained it with a human tendency to narrow their circle of friends with whom they usually communicate. In other words, while people are getting to know each other, they discover their preferences for interlocutors to talk to. It is opposite to the case of the early stage of groups formation, when people communicate more  or less randomly. It is clearly, see in Fig.\ref{fig:results}B, for the NetSense dataset which contains text communication of freshman students who start their studies at a new university. Similarly, we can observe decreasing entropy in other datasets independently of trend of random sequence entropy.

Another observation is that the distance from entropy of the real sequence to entropy of the random sequence, in general, increases over time, see Fig.\ref{fig:entr_comp_zscore}C with the sample of Z-score distance values -- similar trends arise for all other datasets. A group of people unfamiliar to each other engages in nearly random interactions which increasingly become non-random as familiarity of people in the group increases with time.

We recognized some potential of entropy-based measures in solving problems like detection of social communities from dynamic data about human activities. Our hypothesis is that entropy is able to distinguish different groups in the event sequence since the groups may have different dynamic profile of interactions (different entropy levels), e.g. within hospital staff members and separately among patients.

It should be noted that we considered events in the sequence to be directed interactions in our experiments. However, in some applications it may more be meaningful to treat events as undirected contacts.


\section*{Methods}

In this section, we present in details all methods, measures and datasets we used in the experiments. 

\subsection*{Temporal network representation}
All experiments are performed on event sequences ($ES$)\cite{holme2005network}, which are lossless representations of temporal social network and the most popular form of traces about human communication \cite{holme2015modern}. Since it is the most atomic representation, it fits to the real processes better than aggregated approaches like an aggregated weighted network\cite{holme2013epidemiologically} or a time-window graph\cite{krings2012effects}.

An event sequence ($ES$) is a time ordered list of events and each event $ev_{ijk}$ captures a single time-stamped interaction between two individuals in the observed system, i.e. $ev_{ijk}$ is a triple $ev_{ijk} = \{s_i,r_j,t_k\}$, where $s_i$ is the sender/initiator and $r_j$ -- the receiver of interaction at time $t_k$. We also assume that the event can happen only between two different individuals (nodes):
\begin{equation}
    \forall {ev_{ijk} = \{s_i,r_j,t_k\}} s_i \neq r_j
    \label{eq_diff_event}
\end{equation}
We also want to define $e_{ij}$ as an edge between two nodes, that is $e_{ij} = (s_i,r_j)$. It exists if there is any event from $s_i$ to $r_j$ at any time. Note that edges are directed: $(s_i,r_j) \ne (r_j,s_i)$, i.e. $e_{ij} \ne e_{ji}$. The set of all edges derived from a given event sequence $ES$ is denoted as$E$. Let us define $V$ as a set of all distinct individuals (nodes) participating in all considered events, i.e. $V=\{s,r:(s,r) \in E \lor (r,s) \in E, s \ne r \}$. $N$ denotes the size of set $V$: $N=|V|$. 
For further consideration let us define the space of possible edges $\Omega(E)$, i.e. the set of all possible pairs $\{ (s,r): s,r \in V, s \ne r \}$. Hence, $|\Omega(E)|=N(N-1)$.

Some measures in the experiments are computed for the aggregated network, which is a static generalization of the event sequence $ES$ that is simply a directed graph $G$ defined by a tuple: $G=(V,E)$. 

\subsection*{Entropy-based measures for temporal network}
\label{sec_new_measures}
In this section, we would like to propose a holistic approach - new measures for temporal networks designed especially to quantify temporal networks properties in terms of inner dynamic processes. The proposed measures are the main novelty of this work, even though they implement entropy -- the concept well known in physics and information theory. Entropy is a probabilistic description of general systems properties capturing its randomness level. In particular, based on the event sequence ($ES$) as the representation of temporal network, we propose various entropy measures.

In general, we utilize entropy $S$ known in information theory as \textit{information entropy} or \textit{Shannon entropy}, which is defined as follows:
\begin{equation}
    S=-\sum_{i \in O}p(i) ln(p(i))
    \label{eq_entropy_information}
\end{equation}
where $p(i)$ is occurrence probability of state or object $i$, and $O$ is the set of all possible states/objects \cite{shannon2001mathematical}.

\subsubsection*{First-order (node) entropy}
The first approach is based on probability of occurrences of individual nodes $s_i,r_j \in V$, i.e. humans participating in interactions -- events $ev_{ijk} = \{s_i,r_j,t_k\}$. It is the first-order entropy measure that can be considered in three variants: 1) node being a speaker/sender $s_i$, 2) node as a listener/receiver $r_j$ or 3) node occurring as a speaker $s_i$ or listener $r_j$. Using the basic definition of entropy (1), we define the first-order (node) entropy $S_1$ as:

\begin{equation}
    S_1 = -\sum_{v \in V} p_1(v) ln(p_1(v))
    \label{eq_node_entropy}
\end{equation}
where $p_1(v)$ is probability of occurrence for node $v \in V$ in the appropriate role -- the sender, receiver or any of these two. Choice of the role (and the entropy version) depends on what kind of analysis we want to perform. In this paper, we use probability of node occurrences as the sender, because we assume that interaction initiators are more significant than the receivers.

The node entropy measures the diversity of node popularity in the temporal network. In other words, the greater entropy means that the nodes have rather equal probability of occurrence and the small one denotes that some nodes occur significantly more frequently than the others. Entropy has the maximum value when probabilities for all nodes from $V$ are equal. The equal probabilities emerge when all nodes occur the same number of times, e.g. only once or all twice etc. Hence, the equal probabilities are:
\begin{equation}
    p_1(1)=p_1(2)=...=p_1(N)=\frac{1}{N}
    \label{eq_nodes_entropy_prob}
\end{equation}

Then, the maximum possible value of entropy for a given set of nodes $V$ is defined as:
\begin{equation}
    S_1^M = -\sum_{v \in V}\frac{1}{N}ln(\frac{1}{N}) = -|V|*\frac{1}{N}ln(\frac{1}{N}) = ln(N)
    \label{eq_nodes_entropy_max}
\end{equation}

\subsubsection*{Second-order (edge) entropy}
The second approach utilizes probabilities of occurrence of edge $e_{ij}$ from $E$.
We defined the second-order (edge) entropy, as:
\begin{equation}
    S_2 = -\sum_{e_{ij} \in E} p_2(e_{ij}) ln(p_2(e_{ij}))
    \label{eq_event_entropy}
\end{equation}
where $p_2(e_{ij})$ is a probability of edge $e_{ij}$, i.e. probability that events $ev_{ijk}$ are related to edge $e_{ij}$. 
This entropy of the temporal network provides information about how uncertain (random) pairs of nodes (individuals) interact with each other. The greater edge entropy value reflects that the distribution of participating pairs is close to uniform distribution while the smaller value means that some pairs interact more frequently than the others.

We can estimate the maximum value of edge entropy, assuming that probabilities for all possible edges are equal, i.e. all possible pairs of nodes $s_i, r_j$ appear in the same number of events $ev_{ijk}$.
The number of possible edges is $|\Omega(E)|=N(N-1)$. Then, we have:
\begin{equation}
    p_2(e_{12})=p_2(e_{13})=...=p(e_{N1})=...=p_2(e_{N(N-1)})= \frac{1}{N(N-1)}
    \label{eq_event_entropy_prob}
\end{equation}
With this probability maximum value of edge entropy would be defined as:
\begin{equation}
    S_2^M = -\sum_{e_{ij} \in \Omega(E)}\frac{1}{N(N-1)} ln(\frac{1}{N(N-1)}) = -N(N-1) * \frac{1}{N(N-1)} ln(\frac{1}{N(N-1)}) = ln(N(N-1))    
    \label{eq_event_entropy_max}
\end{equation}

For larger number of nodes (large $N$), we have: $S_2^M \approx 2S_1^M$.

\subsubsection*{Third-order (succession) entropy}
The next approach is based on probability of occurrence two particular node pairs (edges) in events one directly after another. We refer to such a pair of edges as succession. Event sequence $ES$ is a list of $M$ events ordered by time: $ES=(ev_1, ev_2, ..., ev_k, ev_{k+1}, ..., ev_M)$, and $ev_k=(s_i,r_j,t_k), ev_{k+1}=(s_{i'},r_{j'},t_{k+1}) \Leftrightarrow t_k \leq t_{k+1}$. For two consecutive events $ev_k$ and $ev_{k+1}$, we can extract participating nodes $s_i,r_j,s_{i'},r_{j'}$, respectively, i.e. edges $e_{ij}, e_{i'j'} \in E$. Such two edges define the single $k$th edge succession occurrence $sc_k=(e_{ij}, e_{i'j'})$ and the set of distinct successions (unique pairs of edges) is denoted by $SC$. Obviously, it may happen that $e_{ij} = e_{i'j'}$. The set of all potentially possible successions is $\Omega(SC)$ with size $|\Omega(SC)|$. This size is limited by the maximum size of the edge set $E$ for a given set of nodes $V$: $|\Omega(SC)| = |\Omega(E)|^2=N^2(N-1)^2$.


Using probability of succession we can define succession entropy:
\begin{equation}
    S_3= -\sum_{sc \in SC} {p_3(sc)ln(p_3(sc))}
    \label{eq_succession_entropy}
\end{equation}
where $p_3(sc)$ is a probability of edge succession $sc$. 

The value of succession entropy quantifies information about how uncertain (random) is presence of particular succession of edge pairs in the event sequence. Similarly to previous approaches, we can find the maximum value of succession entropy by assuming equal distribution of succession probabilities:. 
\begin{equation}
    p_3(sc_1)=p_3(sc_2)=p_3(3)= ... = p_3(|\Omega(SC)|) = \frac{1}{|\Omega(SC)|}
    \label{eq_succession_entropy_prob}
\end{equation}
For these probabilities, the maximum value of succession entropy would be:
\begin{equation}
    \begin{split}
        S_3^M = -\sum_{sc \in \Omega(SC)}\frac{1}{|\Omega(SC)|}ln(\frac{1}{|\Omega(SC)|}) = -|\Omega(SC)|\frac{1}{|\Omega(SC)|}ln(\frac{1}{|\Omega(SC)|}) = ln(|\Omega(SC)|) = ln(|\Omega(E)|^2) =\\
         = 2ln(N(N-1)) = 2S_2^M
    \end{split}
    \label{eq_succession_entropy_max}
\end{equation}

For larger quantity of nodes (big $N$): $S_3^M \approx 4S_1^M$.

\subsubsection*{Normalized entropy}
To compare entropies among datasets with different sizes, we propose the normalized entropy for each previously defined entropy. The normalized entropy for event sequence $ES$ is a ratio of regular entropy to its maximum value:
\begin{equation}
S_o^N = \frac{S_o}{S_o^M}
\label{eq_normalized_entropy}
\end{equation} 
where $o$ is one of entropy types; first-order: $o=1$, second-order: $o=2$, or third-order: $o=3$.

Such normalized definition makes it possible to compare entropy for event sequences independent of their sizes -- numbers of participating nodes, i.e. humans in the social network, see Fig. \ref{fig:entr_comp_zscore}A and \ref{fig:dataset_comp}. 

Note that the experiments were carried out in the incremental setup, i.e. maximum entropy $S_o^M$ was re-calculated after each event for the given incrementally (cumulatively) increased event sequence. It means that the number of all participating nodes $N$ increases over time since new nodes appeared in the sequence, see Supplementary Fig. \ref{fig:normalization_factor}B. The value of $N$ directly impacts on maximum entropy value, Eq. \ref{eq_nodes_entropy_max}, \ref{eq_event_entropy_max}, \ref{eq_succession_entropy_max} and as a result on its normalized  version.

\subsection*{Datasets}
All our experiments were carried out on empirical temporal social networks - event sequences - as well as on artificial ones, randomly generated.
\subsubsection*{Real event sequences}
\begin{itemize}
    \item \textbf{NetSense - text messages}. The dataset contains phone and text communication among students at University of Notre Dame. The dataset was created to map peers' social network and contains data from 3 years (6 semesters) starting from September 6, 2011. \cite{striegel2013lessons}
    \item \textbf{Hospital ward dynamic contact network}. This dataset contains the temporal network of contacts between patients, patients and health-care workers (HCWs) and among HCWs in a hospital ward in Lyon, France, from Monday, December 6, 2010 to Friday, December 10, 2010. The study included 46 HCWs and 29 patients. \cite{10.1371/journal.pone.0073970} Contacts were collected using proximity sensors which do not provide direction of the contact. However, for our experiments, we consider it as directed communication for easier comparison with other datasets.
    \item \textbf{Hypertext 2009 dynamic contact network}. The dataset was collected during the ACM Hypertext 2009 conference, where the SocioPatterns project deployed with the Live Social Semantics application. Conference attendees volunteered to wear radio badges that monitored their face-to-face proximity. The dataset published here represents the dynamical network of face-to-face proximity of ~110 conference attendees over about 2.5 days. \cite{Isella:2011qo} Collecting method does not provide direction of contacts but for easier comparison with other datasets, we consider contacts as directed.
    \item \textbf{Manufacturing emails}. This is the internal email communication between employees of a mid-sized manufacturing company. The network is directed and nodes represent employees while events correspond to individual emails. \cite{konect:2016:radoslaw_email}
\end{itemize}

The dataset profiles are presented in Table \ref{tb_datasets}.

\begin{table}[ht]
\centering
\caption{Datasets in numbers}
\label{tb_datasets}
\begin{tabular}{lcc}
Dataset                         & \# of events      & \# of individuals \\
HyperText conference -- meetings			& 20,818					& 113 \\
Netsense -- text messages		& 32,311					& 212 \\
Manufacturing company -- emails            & 82,927                 & 167 \\
Hospital ward dynamic contact network & 32,424           & 75
\end{tabular}
\end{table}

\subsubsection*{Random event sequences}
For each real event sequence, we generate corresponding random event sequences to provide a baseline for our experiments. The new event sequences were generated preserving timestamps and set of nodes from the real event sequence. Hence, the acquired event sequences are the same in size and have the same set of nodes but different distribution and order of events. We generated an event sequence with following algorithm: 
\begin{enumerate}
\item Take the real event sequence $ES$ and extract distinct nodes from event's senders and receivers -- create set of nodes $V$.
\item Take the next event from the real event sequence, starting from the first one and keep its timestamp $t_k$.
\item Randomly select the sender $s_i \in V$ (according to selected distribution).
\item Randomly select the receiver $r_j \in V$ (according to selected distribution).
\item If the sender and receiver are the same, repeat step 4.
\item Create event $ev_{ijk}=(s_i,r_j,t_k)$.
\item If it is the last event in the real sequence $ES$ -- stop, otherwise go to step 2.
\end{enumerate}
We tested the following random selections: with uniform, normal, and exponential distribution. The results of the experiments show that the differences between distributions in terms of entropy are not significant, hence, we have used only the uniform distribution for random generation. 

For each real event sequence, we generated 100 random event sequences.

\subsubsection*{Evaluation}
We used Z-score measure to evaluate distance between entropy value of the real network and its random analogues, see Fig. \ref{fig:entr_comp_zscore}C. The Z-score value is defined as follows:
\begin{equation}
Z = \frac{(S-\mu)}{\sigma}
\label{eq_zscore}
\end{equation}
where $S$ is the observation from the real data and $\mu,\sigma$ are mean and standard deviation of random variable, respectively. In our case, observation $S$ is the value of appropriate entropy ($S_1, S_2, S_3$) for the real event sequence. Randomly generated 100 event sequences, in turn, are aggregated with mean $\mu$ and standard deviation $\sigma$ of their entropy values.

\bibliography{bibliography}

\section*{Acknowledgements}
This work was partially supported by the National Science Centre, Poland, project no. 2016/21/B/ST6/01463; European Union’s Horizon 2020 research and innovation program under the Marie Skłodowska-Curie grant agreement No. 691152 (RENOIR); the Polish Ministry of Science and Higher Education fund for supporting internationally co-financed projects in 2016-2019 no. 3628/H2020/2016/2; The Army Research Laboratory under Cooperative Agreement no. W911NF-09-2-0053 (the Network Science CTA) and by the Office of Naval Research (ONR) Grant no. N00014-15-1- 2640.

\section*{Supplementary Information}
\begin{figure}[ht]
\centering
\includegraphics[width=\linewidth]{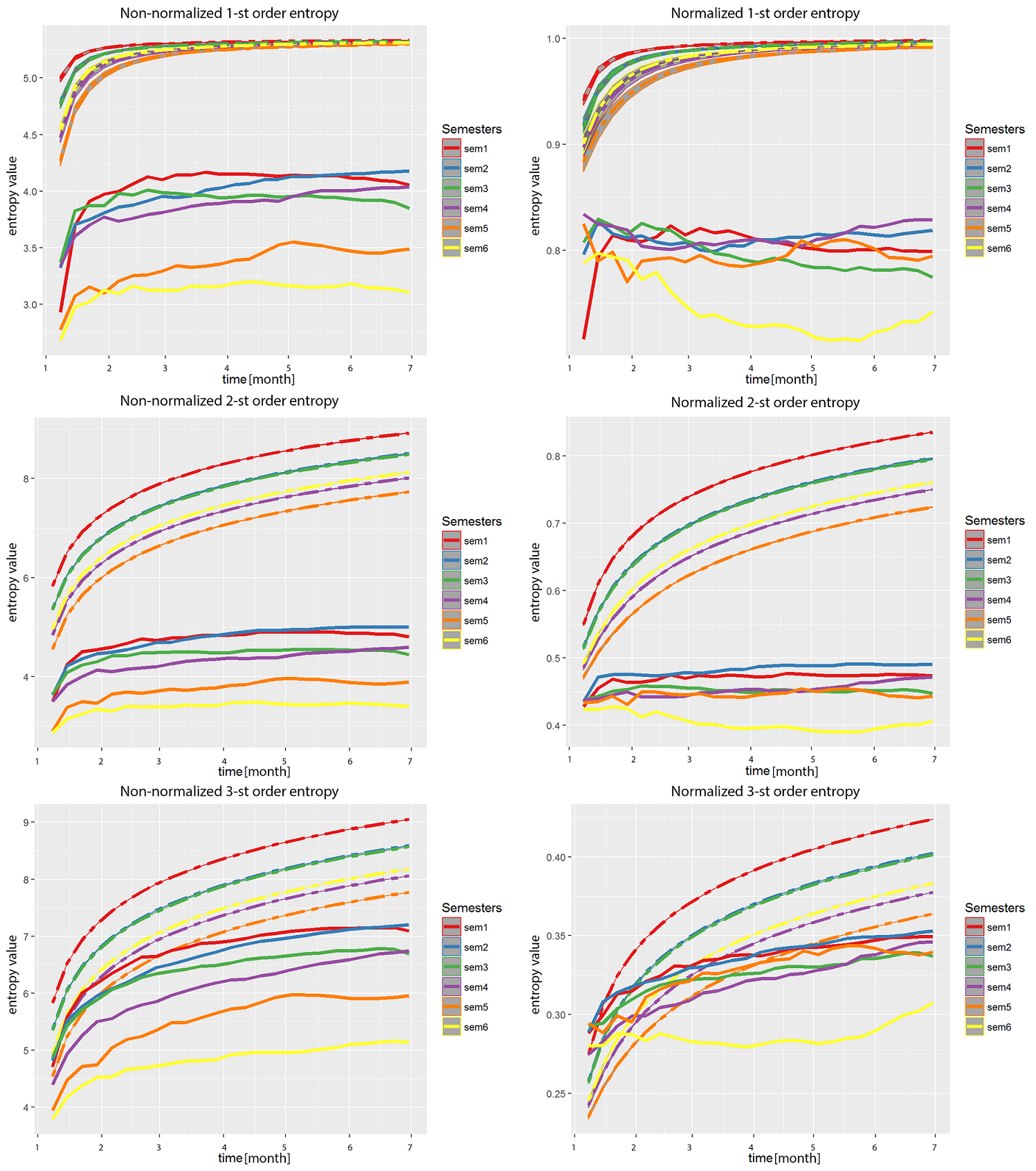}
\caption{All types of entropy for the NetSense dataset. } 
\label{fig:netsense_all}
\end{figure}

\begin{figure}[ht]
\centering
\includegraphics[width=\linewidth]{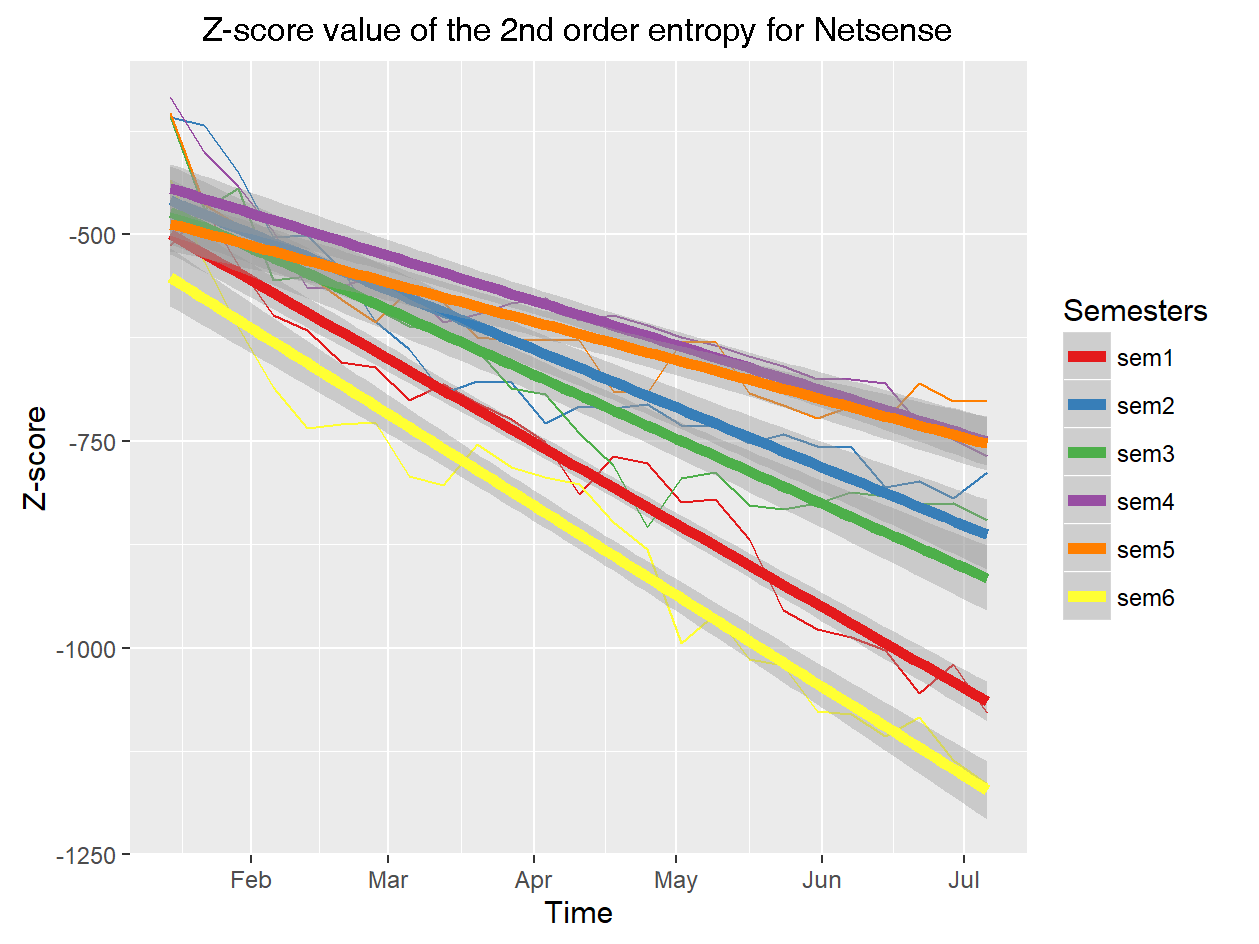}
\caption{Z-score for individual semesters in the NetSense dataset computed for the second-order entropy. Thin lines refer to Z-score values and thick ones with shades are trend lines with standard deviations. Each semester of students communication shows same trend of increasing absolute value of Z-score. In other words, distinction of real data entropy from artificial data entropy become more clear over time.} 
\label{fig:zscore_multi}
\end{figure}

\begin{figure}[ht]
\centering
\includegraphics[width=\linewidth]{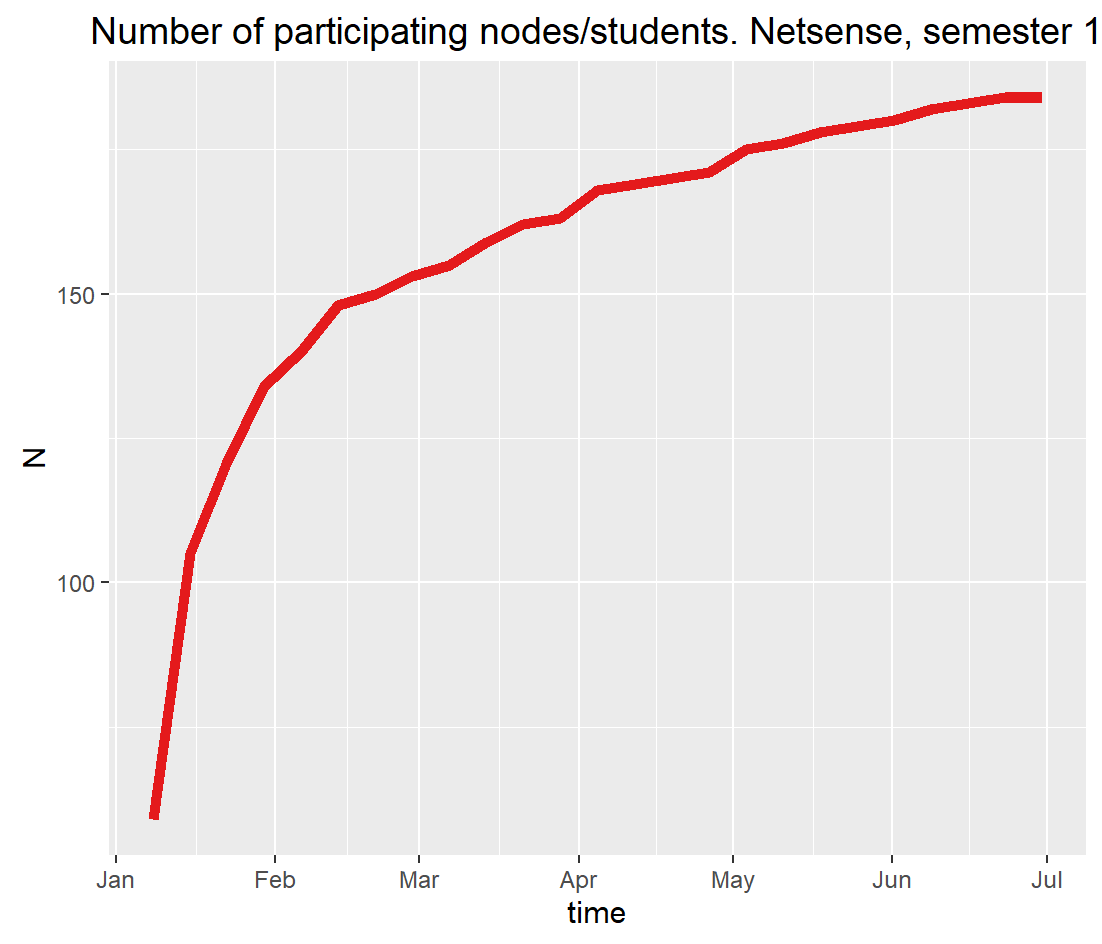}
\caption{NetSense data: the number $N$ of participating nodes -- interacting students -- which non-decrease over time due to incrementing set of events. This is the direct reason for the non-decreasing maximum value of entropy used for normalization.} 
\label{fig:normalization_factor}
\end{figure}

In this section, we present additional experimental results. All entropies computed for the NetSense dataset are presented in Fig.\ref{fig:netsense_all}: the first-, second- and third-order entropies together withe their normalized versions. Z-score values of the second-order entropy for different parts (semesters) of the NetSense dataset are depicted in Fig.\ref{fig:zscore_multi}, while the number of participating nodes (value of $N$) in relation to time is shown in Fig.\ref{fig:normalization_factor}.

\end{document}